\documentclass[letter]{aa}  

\usepackage{graphicx}
\usepackage{txfonts}

\begin{document}

%   \title{Red supergiant surface variability and Gaia parallax uncertainty}
 \title{Probing Red Supergiant dynamics through photo-center displacements measured by Gaia}
 \titlerunning{RSG and Gaia parallaxes}
   \author{A. Chiavassa
          \inst{1,2,3}
          R. Kudritzki
          \inst{4,5}
          B. Davies
          \inst{6}
          B. Freytag
          \inst{7}
          S. E. de Mink
          \inst{3,8,9}
          }

 \institute{Universit\'e C\^ote d'Azur, Observatoire de la C\^ote d'Azur, CNRS, Lagrange, CS 34229, Nice,  France \\
		\email{andrea.chiavassa@oca.eu}
               \and
               Exzellenzcluster 'Origins', Boltzmannstr. 2, 85748 Garching, Germany
               \and
               Max Planck Institute for Astrophysics, Karl-Schwarzschild-Str. 1, 85748 Garching, Germany 
               \and
               LMU M\"{u}nchen, Universit\"{a}tssternwarte, Scheinerstr. 1, 81679 M\"{u}nchen, Germany 
               \and
               Institute for Astronomy, University of Hawaii at Manoa, 2680 Woodlawn Drive, Honolulu, HI 96822, USA 
               \and
               Astrophysics Research Institute, Liverpool John Moores University, Liverpool Science Park ic2, mv 146 Brownlow Hill, Liverpool, L3 5RF, UK 
               \and
         Department of Physics and Astronomy at Uppsala University, Regementsv\"agen 1, Box 516, SE-75120 Uppsala, Sweden
               \and
               Anton Pannekoek Institute of Astronomy and GRAPPA, Science Park 904, University of Amsterdam, 1098XH Amsterdam, The Netherlands
               \and
               Center for Astrophysics $|$ Harvard $\&$ Smithsonian, 60 Garden St., Cambridge, MA 02138, USA
             }

   \date{...}

 \abstract
% 5 {} token are mandatory
 {Red supergiant (RSGs) are cool massive stars in a late phase of
     their evolution when the stellar envelope becomes fully
     convective. They are the brightest stars in the universe at
       infrared light and can be detected in galaxies far beyond the
     Local Group, allowing for accurate determination of chemical
     composition of galaxies. The study of their physical properties is extremely important for various phenomena including the final fate of massive stars as type II supernovae and gravitational wave progenitors.}
          {We explore the well-studied nearby young stellar cluster $\chi$
     Per, which contains a relatively large population of RSG
     stars. Using Gaia EDR3 data, we find the distance of
     the cluster (d = 2.260$\pm$0.020 kpc) from blue main
       sequence stars and compare with RSG parallax measurements
     analysing the parallax uncertainties of both groups. We then
     investigate the variability of the convection-related surface
     structure as a source for parallax measurement uncertainty.}
     {We use state-of-the-art three-dimensional radiative hydrodynamics simulations of convection with
 CO5BOLD and the post-processing radiative transfer code OPTIM3D to compute intensity maps in the Gaia $G$ photometric system. We calculate the variabiltiy, as a function of time, of the intensity-weighted mean (or the photo-center) from the synthetic maps. We then select the RSG stars in the cluster and compare their uncertainty on parallaxes to the predictions of photocentre displacements.}
 {The synthetic maps of RSG show extremely irregular and temporal variable surfaces due to convection-related dynamics. Consequentially, the position of the photo-center varies during Gaia measurements between 0.033 and 0.130~AU ($\approx$1 to $\approx$5$\%$ of the corresponding simulation stellar radius). We argue that the variability of the convection-related surface structures accounts for a substantial part of the Gaia EDR3 parallax error of the RSG sample of $\chi$ Per.}
 {We suggest that the variation of the uncertainty on Gaia  parallax could be exploited quantitatively using appropriate RHD simulations to extract, in a unique way, important information about the stellar dynamics and parameters of RSG stars.}

   \keywords{stars: atmospheres --
                stars: RSG --
                astrometry --
                parallaxes --
                hydrodynamics --
                convection}
   \maketitle
%
%-------------------------------------------------------------------

\section{Introduction}
Red supergiants (RSG) are cool massive stars in a late phase of their
evolution when the stellar envelope becomes fully convective. They are
brightest stars in the universe at infrared light. They can be easily detected as individual stellar
objects in galaxies far beyond the Local Group, where they provide
unique information about chemical composition and galaxy evolution
through the quantitative spectral analysis of their infrared spectra \citep{2015ApJ...805..182G}. In super star clusters (SSC), although small in
numbers,  they dominate the infrared SEDs \citep{2014ApJ...787..142G} allowing for accurate determination of chemical composition of galaxies out to
20 Mpc \citep{2017ApJ...847..112D,2015ApJ...812..160L}. At the same time, RSG are the direct progenitors of type II supernovae and as such crucial
components of galaxies \citep{2018MNRAS.474.2116D}. RSG also play a key role in formation channels for gravitational wave sources through common envelope evolution \citep[e.g.][] {Belczynski+2016, Klencki+2021}. It is thus imperative to investigate the physical
properties of these important sources of astrophysical information in
as much detail as possible.

A crucial step is the determination of RSG luminosities based on the
accurate measurement of distances. For Milky Way RSG the use of
Gaia parallaxes seems ideal for this purpose. Gaia
\citep{2016A&A...595A...1G} is an astrometric, photometric, and
spectroscopic space mission performing a whole sky survey including a
large part of the Milky Way. The most recent release \citep[Gaia Early Data Release 3, ][]{2021A&A...649A...1G} provides parallaxes of
unprecedented accuracy.

However, the situation is not as simple. It is complicated by the
intrinsic variability of RSGs, which to a large extend is caused by
convection related processes in the envelope and at the
surface. In the context of Gaia astrometric measurements, this
convection-related variability, can be considered as a source of "noise" that
needs to be quantified to better characterize any resulting error on
the parallax determination. Most importantly, the motion of convective
cells leads to the surface brightness distribution
over the stellar surface. These dynamical processes can thus manifest themselves as an apparent change of the position of the star as the photo-center (defined as the intensity weighted geometric mean) moves across the stellar surface. This, in turn, can affect the measurement of parallaxes. The first observational evidence for this effect came from measurements with the Hipparcos satellite, where acceptable fits to the Betelgeuse and Antares (two RSG proto-types) astrometric data could not be found \cite{2007A&A...474..653V} and some supplementary noise had to be added to yield acceptable solutions. \cite{2008AJ....135.1430H} conjectured that photo-center motions might be the cause and \cite{2011A&A...528A.120C}, before the Gaia launch, proposed that large-scale convective motions in the photo-center should account for a substantial part of the Hipparcos cosmic noise. More recently, the convection signature in astrometric data of Asymptotic Giant Branch (AGB) stars had been also shown by \cite{2018A&A...617L...1C} and later confirmed by
interferometric images 
\citep{2020A&A...640A..23C}.\\
However, while the photo-center variability appears as a
stumbling block for those primarily interested in obtaining accurate astrometry, it also provides a unique opportunity.
Information about stellar properties, such as the fundamental
parameters and convection properties can be extracted from the
standard deviations of Gaia parallax measurements by using
appropriate radiation-hydrodynamics (RHD) simulations of stellar convection.

The purpose of the work presented here is to demonstrate the feasibility of this. We focus on the nearby
young cluster $\chi$~Per, for which accurate parallaxes are measured
for its bright blue main sequence stars and which also contains a
relatively large number of RSG with Gaia EDR3 parallaxes. We
will compare parallax standard deviations of both groups and explore
the effect of convection-related surface structures on the
photo-center to estimate its impact on the Gaia astrometric
measurements.

\section{Gaia EDR3 cluster $\chi$~Per data: parallaxes and measurement uncertainty}

The nearby young and well-studied cluster $\chi$~Per has a relatively large
population of RSGs \cite[][ and references
therein]{2010ApJS..186..191C}. \cite{2019MNRAS.486L..10D} have used {{\sc
    Gaia}} DR2 data to determine a distance of d = 2.25$\pm$0.15
kpc. With the improvements obtained by Gaia EDR3 we can now
repeat the distance determination and concentrate on a comparison of
main sequence stars and RSG parallaxes.

\begin{figure*}[!h]
   \centering
    \begin{tabular}{cc}

    \includegraphics[width=0.34\hsize]{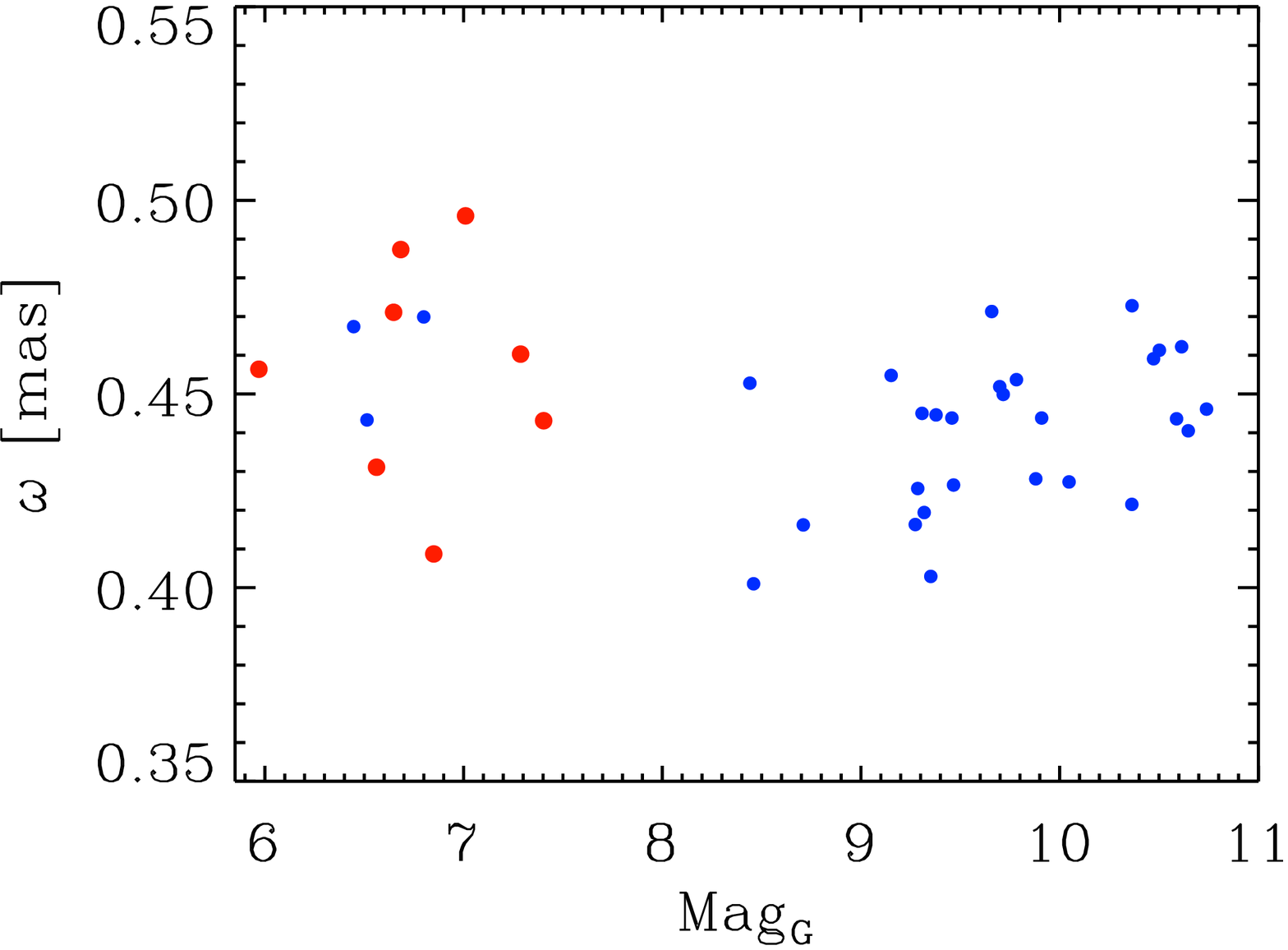} 
        \hspace{2.cm}
     \includegraphics[width=0.35\hsize]{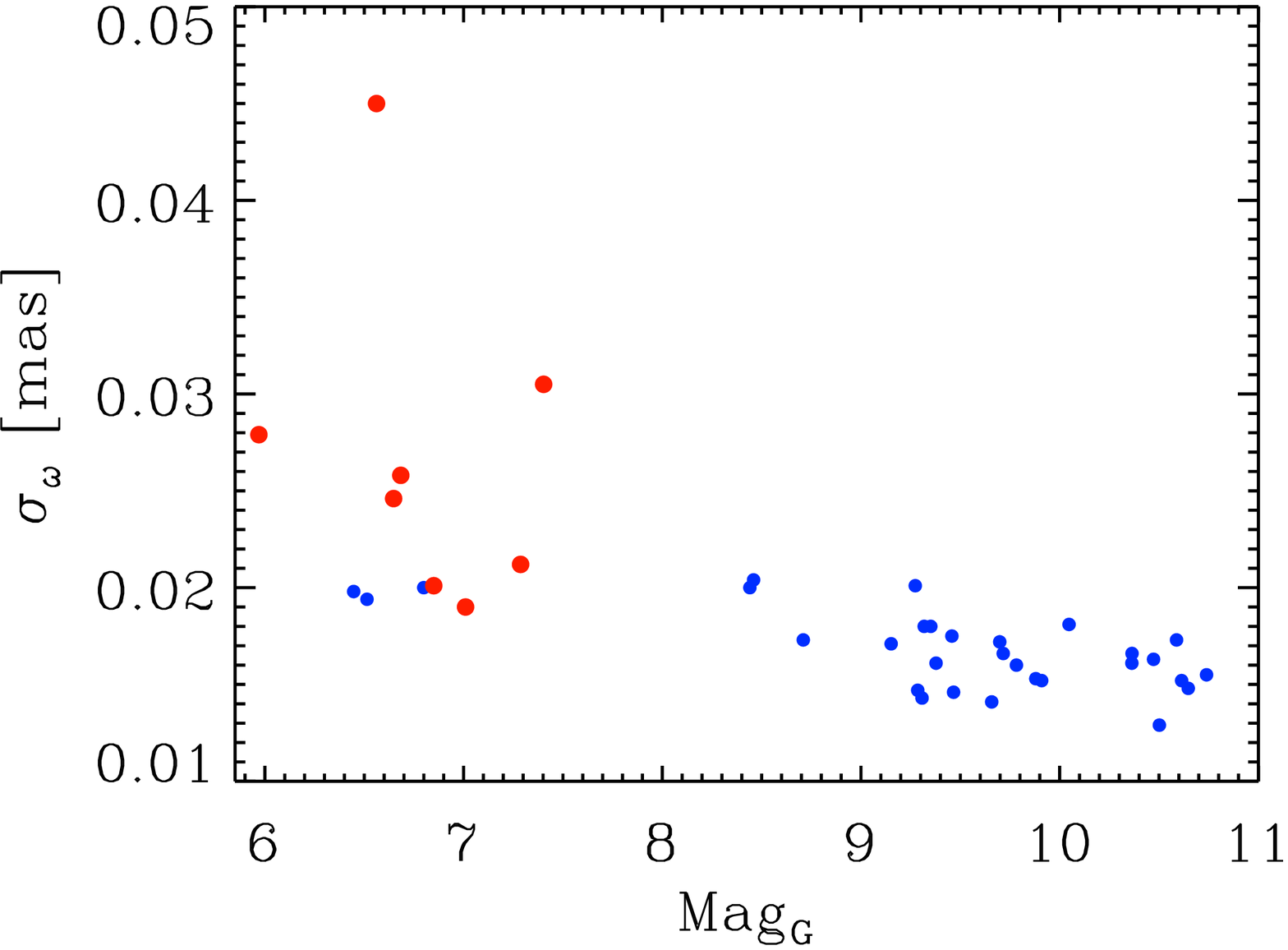} 
        \vspace{0.2cm}
  \end{tabular}
      \caption{Gaia EDR3 data for parallax ($\varpi$,
        \emph{left panel}) and the measurement uncertainties
        ($\sigma_{\varpi}$, \emph{right panel}) for the stars of
        Table~\ref{observations}. The x-axis displays the Magnitude in the Gaia $G$ photometric system. Highlighed in red the RSG stars,
        while the sample of blue main sequence stars (see text) is
        shown in blue.}
        \label{im1}
           \end{figure*}

In a first step, we focus on bright main sequence stars with Gaia EDR3
magnitude $G$ $\le$ 10.8 mag and effective wavenumber $\nu_{\rm{eff}} \ge$
1.5 $\mu^{-1}$. Following \cite{2019MNRAS.486L..10D} we
use Gaia EDR3 proper motions as a criterium for cluster
membership. We regard all objects with

\begin{equation}
  {(p_{\alpha} - p_{\alpha}^{c})^2 \over \sigma_{\alpha}^2} +
  {(p_{\delta} - p_{\delta}^{c})^2 \over \sigma_{\delta}^2} \le 1
\end{equation}

as cluster members. Here, p$_{\alpha}$ and p$_{\delta}$ are the Gaia
EDR3 proper motions in right ascension and declination, respectively,
measured in mas\,yr$^{-1}$. p$_{\alpha}^{c}$ = -0.64 mas\,yr$^{-1}$ and
p$_{\delta}^{c}$ = -1.17 mas\,yr$^{-1}$ are the central values for the
sample. $\sigma_{\alpha}$ = 0.303 mas\,yr$^{-1}$ and $\sigma_{\delta}$ = 0.222
mas\,yr$^{-1}$ define the borderline for membership. Fig.~\ref{im1} shows
parallaxes ($\varpi$) of the sample selected in this way and their
uncertainty ($\sigma_{\varpi}$). We note that we have applied a zero-point
correction to the parallaxes following \cite{2021A&A...649A...4L}, see
their equations A3, A4, A5 and Table 9. The zero-point correction
magnitude dependence is the reason for the restriction to $G\le10.8$
mag of our sample. The data for
Fig.~\ref{im1} are given in Table~\ref{observations}. The mean value of the
parallaxes of is $\bar{\varpi}$ = 0.442$\pm$0.004 mas corresponding to
a distance of d = 2.260$\pm$0.020 kpc.

In the second step we select RSG stars in the $\chi$ Per cluster with proper motions in the
same domain as the blue stars selected in previous step. We identify
eight objects. Their parallaxes, G-Band magnitudes and luminosities
are also given in Table A1. We have also applied the parallax zero point
correction to these objects. The luminosities are taken from \cite{2018MNRAS.474.2116D}, but have been corrected for the new
distance to $\chi$ Per obtained from the blue objects.

We include the RSG in Fig.~\ref{im1} and find good agreement with
respect to parallaxes. The RSG mean value is slightly higher,
$\bar{\varpi}_{RSG}$ = 0.457$\pm$0.010 mas, but agrees within the
error margins. However, the measurement uncertainties for the RSG
sample are notably higher than those of the blue star sample, even though it can be noted that 6 blue points have also higher values ($\sigma_{\varpi}> 0.019$ mas). These objects may be binary systems with an impact on the parallax measurement and this will be explored in Gaia DR3 release. 
In the following, we argue that the high  uncertainties measurement for the RSG is the result of the photo-center variability induced by RSG dynamics. 

\section{Radiation-hydrodynamics simulations to explain the Gaia measurement uncertainty}

We use the RHD code CO$^5$BOLD \citep{2012JCoPh.231..919F}  to compute simulations for RSG stars (Table~\ref{simus}). The code solves the coupled non-linear equations of compressible hydrodynamics and non-local radiative energy transfer in the presence of a fixed external spherically symmetric gravitational field in a three-dimensional cartesian grid. Solar abundances are assumed. 

We followed the approach by \cite{2018A&A...617L...1C} and computed intensity maps in the Gaia $G$ photometric system \citep{2021A&A...649A...3R},
using the radiative transfer OPTIM3D-code \citep{2009A&A...506.1351C} for all the snapshots from the RHD simulations. This code takes into account the Doppler shifts caused by the convective motions. The radiative transfer is computed in detail using pre-tabulated extinction coefficients from MARCS stellar atmosphere code \citep{2008A&A...486..951G} and for a solar composition \citep{2009ARA&A..47..481A}.

   \begin{figure}[!h]
   \centering
    \begin{tabular}{c}
    \includegraphics[width=1.\hsize]{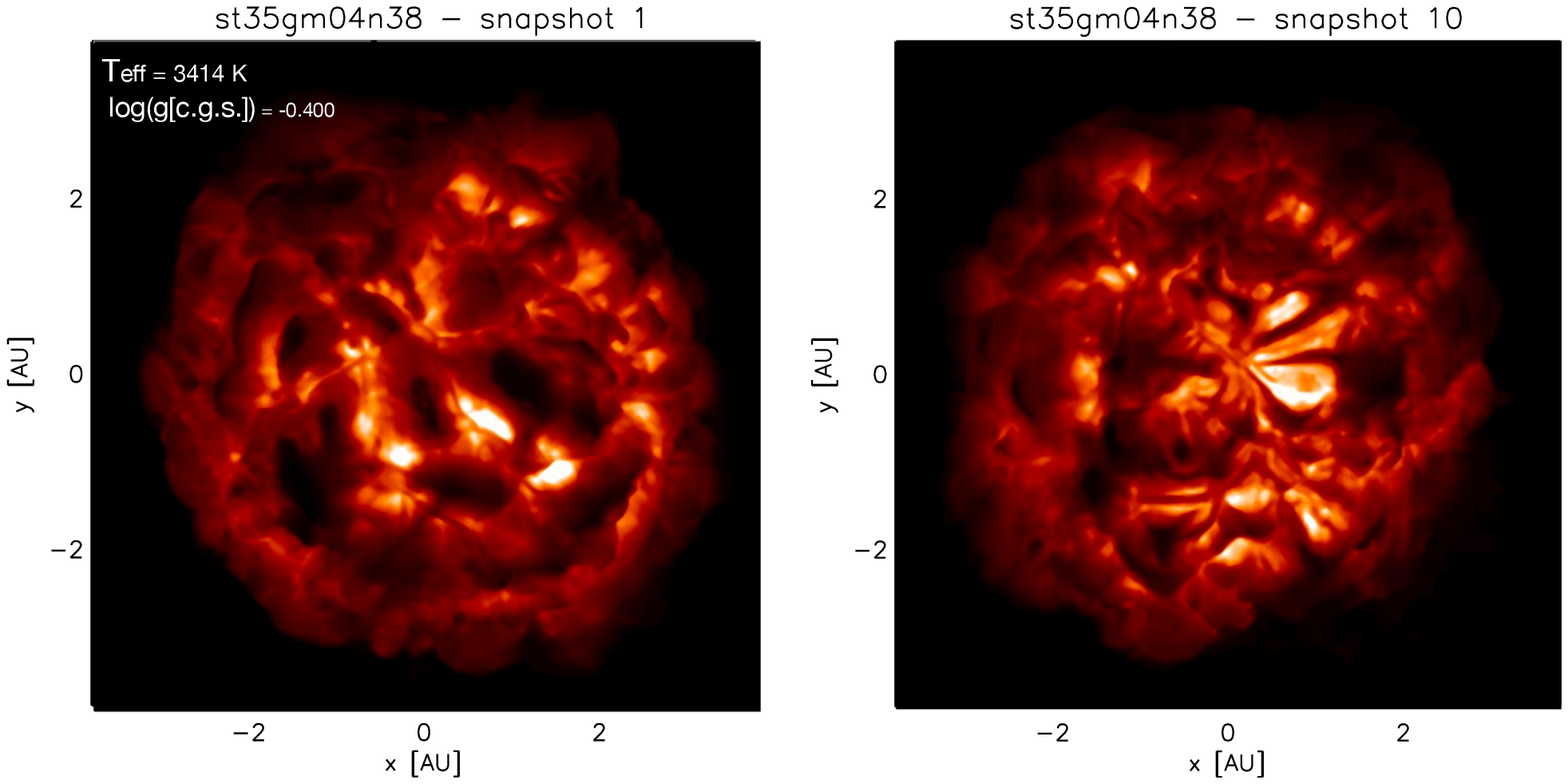}  \\
    \includegraphics[width=1.\hsize]{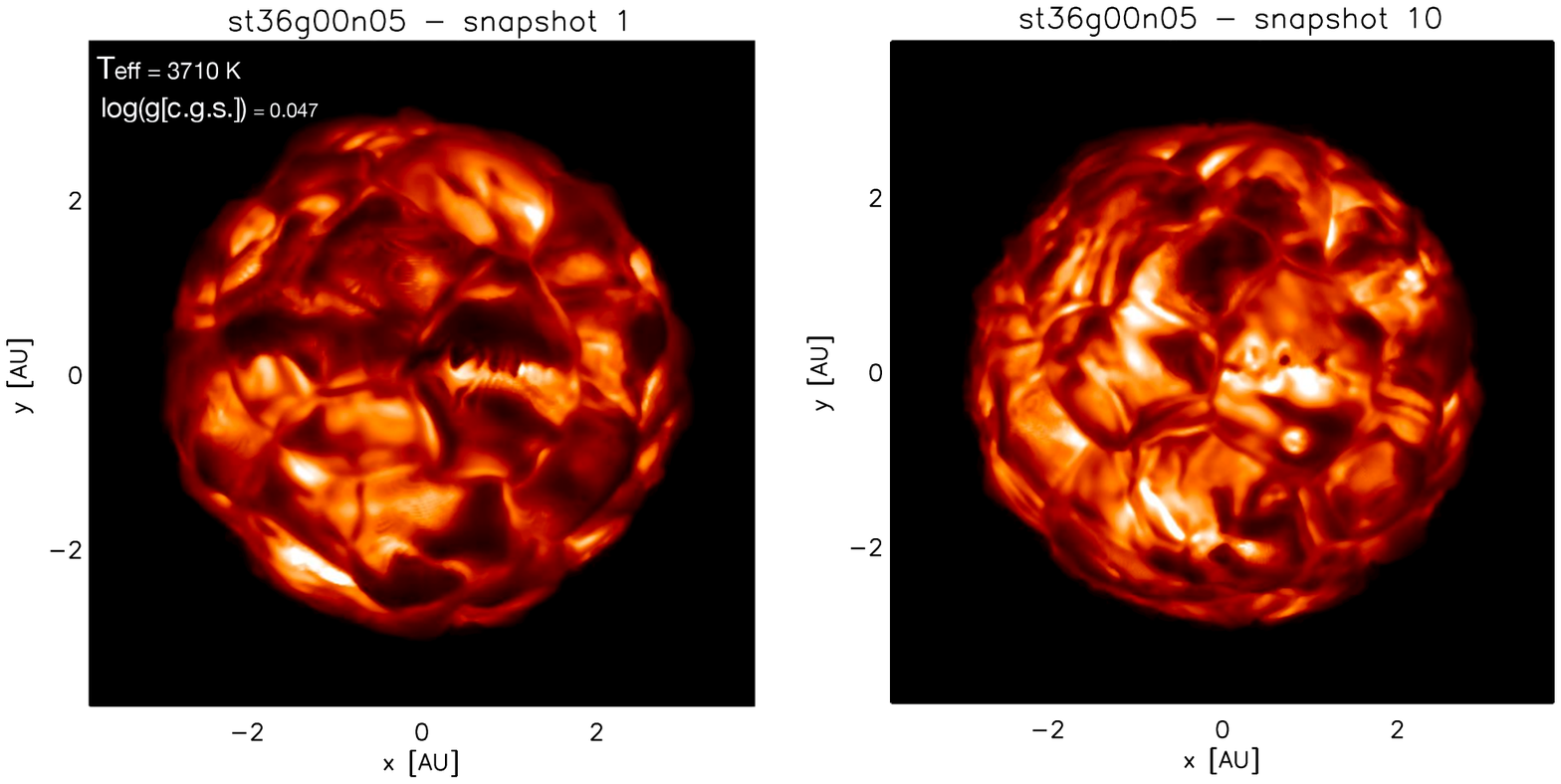}  \\
  \end{tabular}
      \caption{Example of intensity maps in Gaia $G$ photometric system
        for two RHD simulations in Table~\ref{simus} and for two
        different snapshots about 200 days apart. The range is
        $[0.-825942]$ 
        erg\,s$^{-1}$\,cm$^{-2}$\,\AA$^{-1}$\,for low $T_\mathrm{eff}$ and $\log g$ simulation st35gm04n38 (\emph{top panel}) and $[0.-299977]$ erg\,s$^{-1}$\,cm$^{-2}$\,\AA$^{-1}$\, for the highest $T_\mathrm{eff}$ and $\log g$ simulation st36gm00n05 (\emph{bottom panel}). The intensity is shown on a square-root scale to make the structures more visible.}
        \label{imRSGs}
           \end{figure}   

These simulations predict very large variations in velocity, density and temperature that produce strong shocks in their extended photosphere that can cause the gas to levitate and thus contribute to mass-loss \citep{2018A&ARv..26....1H,2017A&A...600A.137F,2011A&A...535A..22C}. The stellar surface is characterised by complicated convection-related structures of sizes close to a third of the stellar radii that evolve on several months to years together with short-lived (weeks to months) small scale ones \citep{2017A&A...600A.137F,2011A&A...528A.120C}. The resulting synthetic images in the $G$ photometric system are strongly affected by this in terms of intensity distribution (Fig.~\ref{imRSGs}). As a consequence, the position of the photo-center is expected to change as a function of time during Gaia measurements, as already pointed out in \cite{2011A&A...528A.120C}. 

We calculated the position of the photo-center for each map (i.e., as a function of time) as the intensity-weighted mean of the $x-y$ positions of all emitting points tiling the visible stellar surface according to

\begin{eqnarray}
Px=\frac{\sum_{i=1}^{N} \sum_{j=1}^{N} I(i,j)*x(i,j)}{\sum_{i=1}^{N} \sum_{j=1}^{N} I(i,j)} \\
Py=\frac{\sum_{i=1}^{N} \sum_{j=1}^{N} I(i,j)*y(i,j)}{\sum_{i=1}^{N} \sum_{j=1}^{N} I(i,j)},
\end{eqnarray}

where $I\left(i,j\right)$ is the emerging intensity for the grid point
$(i,j)$ with coordinates $x(i,j)$, $y(i,j)$ of the simulation, and $N$
is the total number of grid points in the simulated box. In presence
of surface brightness asymmetries the photo-center position will rarely
coincide with the barycenter of the star and its position will change
as the surface pattern changes with time. This is displayed in the
photo-center excursion plots for each simulation in
Fig.~\ref{photo_appendix}\footnote{The related videos are visible here: https://doi.org/10.5281/zenodo.6363011}. The averages over time of
photo-center position and its standard deviation $\sigma_P$ are overplotted as the
  central red dot and the red circle, respectively. The coordinates of
  the red dot, $\langle P_x\rangle$ and $\langle P_y\rangle$, are reported in
Table~\ref{simus} together with $\sigma_P$.
$\langle P_x\rangle$ and $\langle P_y\rangle$ are
mostly affected by short time scales corresponding to the small
atmospheric structures but they are significant different from zero, revealing
that the photo-centers typically do not coincide with the nominal center of
the star (dashed lines in Fig.~\ref{photo_appendix}) because of the
presence of convection-related surface structure evolving with
time. On the other hand, $\sigma_P$ varies between 0.033 and 0.130 AU
($\approx$1 to $\approx$5$\%$ of the corresponding stellar
radius). Moreover, $\sigma_P$ correlate with the stellar surface
gravity, that governs the size of granules which, in turn, controls
the photometric variations.% Lower surface gravity (i.e. more extended atmospheres) return larger excursions of the photo-center $\sigma_P$. This behaviour is explained by the correlation between the stellar atmospheric pressure scale height ($H_{\rm{p}}\approx$ T$_{\rm eff}$/g) and the photo-center displacement \citep{2018A&A...617L...1C,2011A&A...528A.120C,2006A&A...445..661L,2001ASPC..223..785F}. 

%It should be noted that the main information that is used to determine the astrometric characteristics of each stars will be the along-scan measurement of Gaia. \cite{2011A&A...528A.120C} showed that projection of the star position along the scanning direction of the satellite with respect to a known reference point disclose similar, though slightly increased, values of $\sigma_P$. At the current state of the DR2, it is not possible to perform this on real data and we assumed the conservative value of $\sigma_P$ directly extracted from the RHD simulations for the following comparisons.

%...............................................................................
% --- Table of model parameters, produced with rhd_printtable.pro ---
\begin{table*}
\small
\begin{center}
 \caption{RHD simulations parameters}
 \label{simus}
 \begin{tabular}{l|rrrrrr|rrr}
\hline
Simulation & $M_\star$ & $\log (L_\star/L_\sun)$ & $R_\star$ & $T_\mathrm{eff}$ & $\log g$ & $t_\mathrm{avg}$ & $\sigma_P$ & $\langle P_x\rangle$ & $\langle P_y\rangle$ \\
& $M_\sun$ &  & AU & K & (cgs) & yr & AU & AU & AU  \\ \hline
%st28gm07n001 & 1.0 & 10028  &  2.48 & 2506 & -1.02 &  30.90 & 2.247 & 1.397 &  0.303  &   0.181 \\
st35gm04b1n001 & 5 &   4.61 &  2.77 & 3373 & $-$0.410 &  24.95   & 0.114 & $-$0.105 &  0.007 \\
st35gm04n38\tablefootmark{a} & 5 &  4.62 &   2.72 & 3414 & $-$0.400 & 11.45  & 0.113 & $-$0.010 &     0.136  \\
st35gm03n13\tablefootmark{b} & 12 &   4.95 &    3.95 & 3430 & $-$0.354 & 9.24  & 0.036 & 0.022 &  0.008 \\
st36gm00n06\tablefootmark{b} & 6 &  4.38 &    1.82  & 3660 & 0.009 & 7.23  & 0.022 & 0.021 &   0.026 \\
st36gm00n04\tablefootmark{b} & 5 &   4.39  &    1.80 & 3663 & 0.023 & 22.92  & 0.030 &0.006 &   0.003  \\ 
st36gm00n05\tablefootmark{b} & 6 & 4.39 &    1.75 & 3710 & 0.047 & 3.75  & 0.031& 0.011 &  $-$0.006 \\
\hline
 \end{tabular}
\end{center}
{The table shows the simulation name, the stellar mass $M_\star$,
  the average emitted luminosity $L_\star$, the average approximate
  stellar radius $R_\star$, effective temperature $T_\mathrm{eff}$,
  and surface gravity $\log g$, and the time $t_\mathrm{avg}$ used for
  the averaging. The last three columns are the standard deviation
    ($\sigma_P$) of the time-averaged values of the photo-center
    displacement and its coordinates $\langle P_x\rangle$ and $\langle P_y\rangle$.}
\tablefoot{\\
\tablefoottext{a}{Simulation presented in \cite{2021arXiv211210695C} and \cite{2019A&A...632A..28K}}\\
\tablefoottext{b}{Simulation presented in \cite{2011A&A...535A..22C}}}
\end{table*}
%...............................................................................

\section{Comparison to observations}

\begin{figure}[!h]
   \centering
    \begin{tabular}{c}
     \includegraphics[width=0.85\hsize]{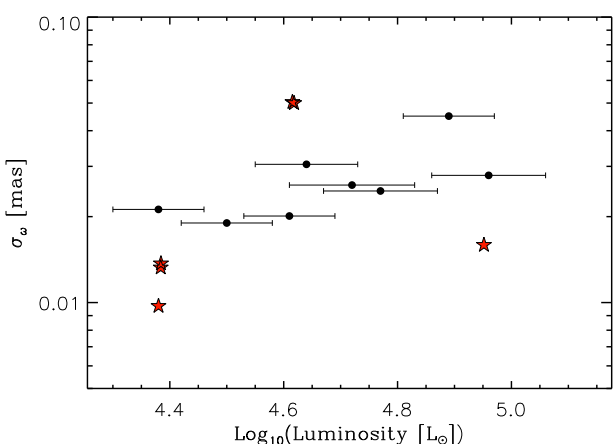} 
  \end{tabular}
      \caption{Absolute luminosity against parallax error
          ($\sigma_{\varpi}$ in Table~\ref{observations}) of the RSG
          stars in $\chi$~Per cluster (black filled circles with error bars)
        compared with the standard errors of the photo-center
        displacements of the RHD simulations (red star symbol). For the
        calculation of the latter we use the standard deviations
        $\sigma_P$ of Table~\ref{simus} and transform to
        $\sigma_{\varpi}$ = $\sigma_P\cdot\bar{\varpi}$ adopting the mean
        parallax of 0.442 mas of the blue star sample (see text).} %The black
%        square displays the position of Betelgeuse if it were in this
 %       cluster. Data for the latter are from \cite{2011A&A...528A.120C} (Table 4) and  \cite{2009AJ....137.3558S}.}}
        \label{imcomparison}
           \end{figure}

In this section we investigate if the parallax errors excess seen for
the RSG stars in $\chi$~Per cluster can be explained by the resulting
motion of the stellar photo-center revealed by the RHD
simulations. For that, Figure~\ref{imcomparison} displays the
comparison between Gaia parallax uncertainty and the standard
deviations of the simulations from Table~\ref{simus}. While none
of our simulations has been computed to exactly represent the stellar parameters of the observed stars, 
the RSGs are within the predictions of the 3D simulations and the
general agreement is good. This attests that convection-related
variability accounts for a substantial part of the parallax error in
Gaia measurements. 

One limitation of this analysis is the
restriction of the 3D grid in stellar parameters. For a better
comparison, one would need extended simulations and observations
  with known luminosities, masses, and radii and spatially resolved
  observations to unveil the presence of convection-related surface
  structures \citep[e.g., ][]{2020A&A...640A..23C}. The latter is
  unfortunately not possible for $\chi$~Per, which is too far in
    distance, however, the evidence for the effects of photo-center
    variability is prominent. 

Given the fact that $\sigma_P$ can explain Gaia measurement uncertainties of the parallaxes, we suggest that parallax variations from Gaia measurements could be exploited quantitatively using appropriate RHD simulations to extract, in a unique way, the fundamental properties of these RSG stars such as the surface gravity that controls the size of the granules and the photometric variations.

\section{Summary and conclusions}

We used Gaia EDR3 measurements of parallaxes and proper
motions of blue main sequence stars and determined the distance of the $\chi$~Per cluster. The mean value of the parallaxes of is $\bar{\varpi}$ = 0.442$\pm$0.004 mas corresponding to a distance of d = 2.260$\pm$0.020 kpc.
We then selected a subset of RSG stars, with proper motions in the
same domain as the blue stars, and find pronounced evidence that the measured Gaia uncertainty of parallaxes is higher than those of the blue star sample.

With the aim of explaining the high uncertainties, we used the snapshots from a grid of RHD simulations of RSG stars to compute intensity maps in the Gaia $G$ photometric system. The synthetic maps show extremely irregular surfaces due to convection-related dynamics. The largest structures evolve on timescales of months/years, while the small ones on timescales of weeks/month. Consequentially, the position of the photo-center is expected to change as a function of time during Gaia measurements. We calculated the standard deviation ($\sigma_P$) of the photo-center excursion for each simulation and found that $\sigma_P$ varies between 0.033 and 0.130~AU ($\approx$1 to $\approx$5$\%$ of the corresponding stellar radius) depending on the simulation. 

We then compared the measurement of Gaia uncertainty on
parallax of the RSG sample to the $\sigma_P$ extracted from the
simulations. The general agreement is good. The predictions of the
  3D simulations enclose the measured RSG observed uncertainty,
albeit these simulations have not been computed to exactly represent the properties of those stars. This suggests that stellar dynamics, quantified through the mean photo-center noise, accounts for a substantial part of the parallax uncertainty for these RSG stars. We suggest that the variation of the uncertainty on Gaia parallax could be exploited quantitatively using appropriate RHD simulations to extract, in a unique way, important information about the stellar dynamics and parameters of RSG stars.
 
\begin{acknowledgements}
This work is funded by the Deutsche Forschungsgemeinschaft (DFG, German Research Fundation) under Germany's Excellent Cluster Strategy $-$ EXC-2094\,$-$\,390783311. AC acknowledges support from the French National Research Agency (ANR) funded project PEPPER (ANR-20-CE31-0002). BF acknowledges funding from the European Research Council (ERC) under the European Union's Horizon 2020 research and innovation programme Grant agreement No. 883867, project EXWINGS) and the Swedish Research Council ({\it Vetenskapsr{\aa}det}, grant number 2019-04059). This work was granted access to the HPC resources of Observatoire de la C\^ote  d'Azur $--$ M\'esocentre SIGAMM and Swedish National Infrastructure for Computing (SNIC) at UPPMAX.
\end{acknowledgements}


\bibliographystyle{aa}
\bibliography{biblio.bib}

%
%-------------------------------------------------------------
%               Appendices have to be placed at the end, after
%                                        \end{thebibliography}
%-------------------------------------------------------------

\begin{appendix} %First appendix

\section{Parallaxes table of the observed stars}

\begin{table*}
\begin{center}
 \caption{Parallaxes and their uncertainty for the $\chi$~Per cluster. The top list displays the RSG stars, while the bottom one the blue main sequence objects.}
 \label{observations}
 \begin{tabular}{cccccc|c}
\hline
$\varpi$\tablefootmark{a}  & $\sigma_{\varpi}$   &  $G$\tablefootmark{a} & $\sigma_{G}$ & $\log (L_\star/L_\sun)$\tablefootmark{b} & $\sigma_{L}$ & Name \\
$[mas]$  & $[mas]$  & & & & & \\
\hline
%0.1690  &   0.1073  &  0.366469&   13.639342 &  0.005116 &     5.31  &   0.10  &  Cl* Westerlund 1 W20 \\
%0.3370  &  0.1235  & 0.3665 &    11.259 &  0.007   &   4.96  &  0.11  &  Cl* Westerlund 1 W237 \\
%0.0893 &  0.1477   &   1.6540& 14.705 &  0.004  &    4.78  &  0.15  &  Cl* Westerlund 1 W75 \\
%0.0534  &  0.0917  &  1.7172&   11.311 &   0.008 &    5.44 &  0.11  &  Cl* Westerlund 1 W26 \\
0.4564 &  0.0279     & 5.970 &  0.005 &      4.96 &  0.10 &  SU Per \\
0.4311 &   0.0450    & 6.562 &  0.007  &     4.89 &  0.08 &  RS Per \\
0.4711 &   0.0246    &  6.648 &  0.004   &   4.77 &   0.10  &  AD Per \\
0.4873 &  0.0258    & 6.684 &  0.004  &   4.72 &  0.11  &  V441 Per \\
0.4431 &   0.0305   &   7.402 &  0.006 &     4.64 &   0.09  &  BU Per \\
0.4087 &  0.0201    &   6.850 &  0.004   &   4.61  &  0.08  &  FZ Per \\
0.4960 &  0.0190   &  7.010 &  0.003   &     4.50  &  0.08  &   V439 Per \\
0.4603 &  0.0212   &   7.287 &  0.003  &    4.38  &  0.08  &  V403 Per \\
%0.3398 &   0.0708   &   0.208358 &  10.224579 &  0.014203 &      5.11 &   0.06 &   MY Cep \\
%0.3530 &  0.0336  &   0.0952 &  10.204 &  0.003  &    4.48 &   0.05  &  Cl* NGC 7419 BMD 435 \\
%0.3155 &  0.0324   &   0.1027 &  10.300 &  0.003  &   4.48 &    0.06  &  Cl* NGC 7419 BMD 696 \\
%0.3374 &  0.0427   &  0.1266 &   10.944 &  0.003  &     4.39  &  0.06  &   Cl* NGC 7419 BMD 139 \\
%0.3442 &  0.0309   &  0.0898 &   10.611 &   0.003  &   4.33  &  0.07  &  Cl* NGC 7419 BMD 921 \\
\hline
\hline
     0.4265 &    0.0146 &         9.466 & 0.002 &  &  &  Cl* NGC 869 W 304  \\
     0.4728 &    0.0166 &         10.364 & 0.002 & &  & Cl* NGC 869 W 300 \\
     0.4436 &    0.0173 &         10.588 & 0.002 & &  & NSV 776\\
     0.4713 &    0.0141 &         9.658 & 0.002 & &  &  Cl* NGC 869 W 288\\
     0.4537 &    0.0160 &         9.782 & 0.003 & &  & BD+56 515 \\
     0.4461 &    0.0155 &         10.739 & 0.003 & &  & BD+56 517 \\
     0.4256 &    0.0147 &         9.286 & 0.003 & &  &  BD+56 518\\
     0.4519 &    0.0172 &         9.698 & 0.003 & &  &  BD+56 519\\
     0.4548 &    0.0171 &         9.152 & 0.003 & &  & V* V614 Per\\
     0.4528 &    0.0200 &         8.441 & 0.003 & &  & BD+56 521\\
     0.4446 &    0.0161 &         9.378 & 0.005 & &  & BD+56 523\\
     0.4450 &    0.0143 &         9.307 & 0.003 & &  & Cl* NGC 869 HG 1085\\
     0.4281 &    0.0153 &         9.880 & 0.003 & &  & BD+56 526\\
     0.4613 &    0.0129 &         10.502 & 0.003 & &  & BD+56 528\\
     0.4194 &    0.0180 &         9.318 & 0.003 & &  & BD+56 529 \\
     0.4438 &    0.0152 &         9.909 & 0.003 & &  & Cl* NGC 869 LAV 1092\\
     0.4215 &    0.0161 &         10.363 & 0.003 & &  & Cl* NGC 869 LAV 1101\\
     0.4405 &    0.0148 &         10.647 & 0.003 & &  & BD+56 533\\
     0.4622 &    0.0152 &         10.614 & 0.003 & &  & HD 14162 \\
     0.4433 &    0.0194 &         6.514 & 0.003 & &  & HD 14210 \\
     0.4273 &    0.0181 &         10.047 & 0.003 & &  &BD+56 548 \\
     0.4010 &    0.0204 &         8.459 & 0.003 & &  & HD 14321\\
     0.4162 &    0.0173 &         8.710 & 0.003 & &  & HD 14357\\
     0.4699 &    0.0200 &         6.800 & 0.003 & &  &BD+56 563 \\
     0.4438 &    0.0175 &         9.458 & 0.003 & &  & Cl* NGC 884 W 222\\
     0.4029 &    0.0180 &         9.351 & 0.003 & &  & Gaia DR2 458454640368346624\\
     0.4591 &    0.0163 &         10.472 & 0.005 & &  & Gaia DR2 458406124415716224\\
     0.4163 &    0.0201 &         9.273 & 0.003 & &  & Cl* NGC 884 W 168\\
     0.4674 &    0.0198 &         6.447 & 0.003 & &  & 2MASS J02220081+5707320\\
     0.4499 &    0.0166 &         9.716 & 0.005 & &  & BD+56 571\\
\hline
\end{tabular}
\end{center}
{}
\tablefoot{\\
\tablefoottext{a}{Gaia EDR3 \citep{2021A&A...649A...1G}}\\
\tablefoottext{b}{\cite{2018MNRAS.474.2116D}}\\
}
\end{table*}

\section{Photo-center position for the different RHD simulations}

\begin{figure*}%f1
    \begin{tabular}{c}
         \vspace{0.7cm}
    \includegraphics[width=0.47\hsize]{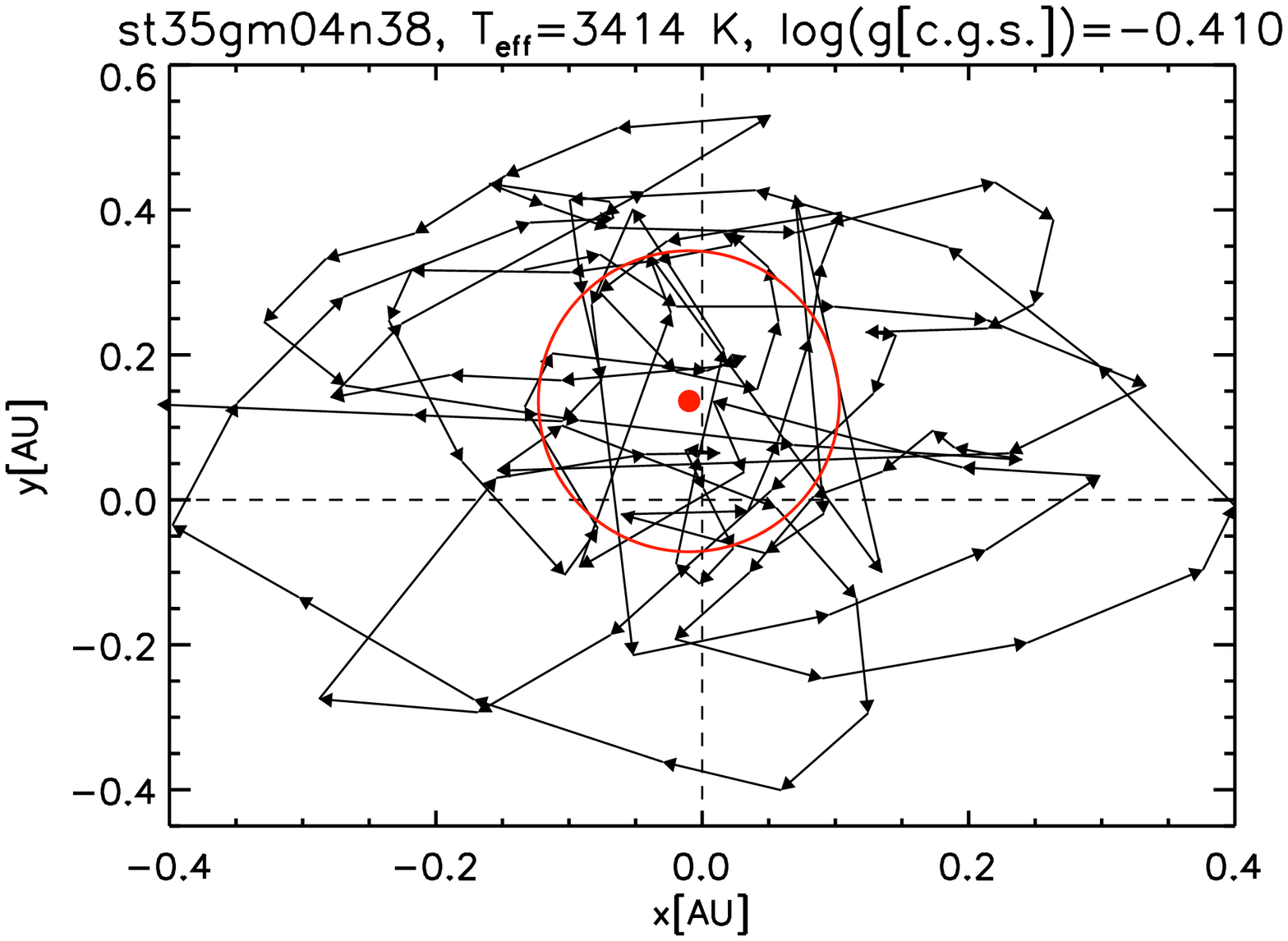} 
    \hspace{0.8cm}
     \includegraphics[width=0.47\hsize]{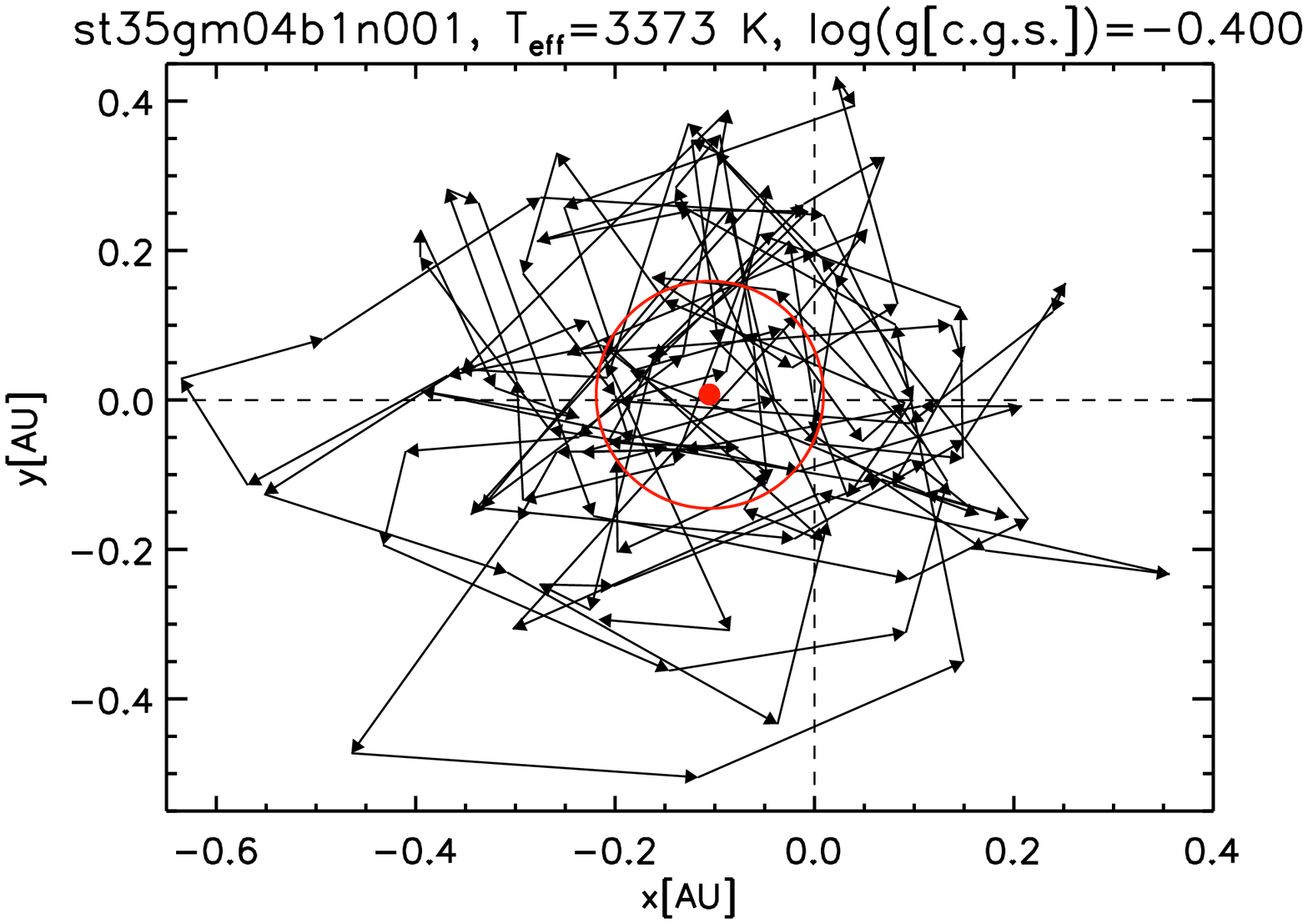} \\
       \vspace{0.7cm}
      \includegraphics[width=0.47\hsize]{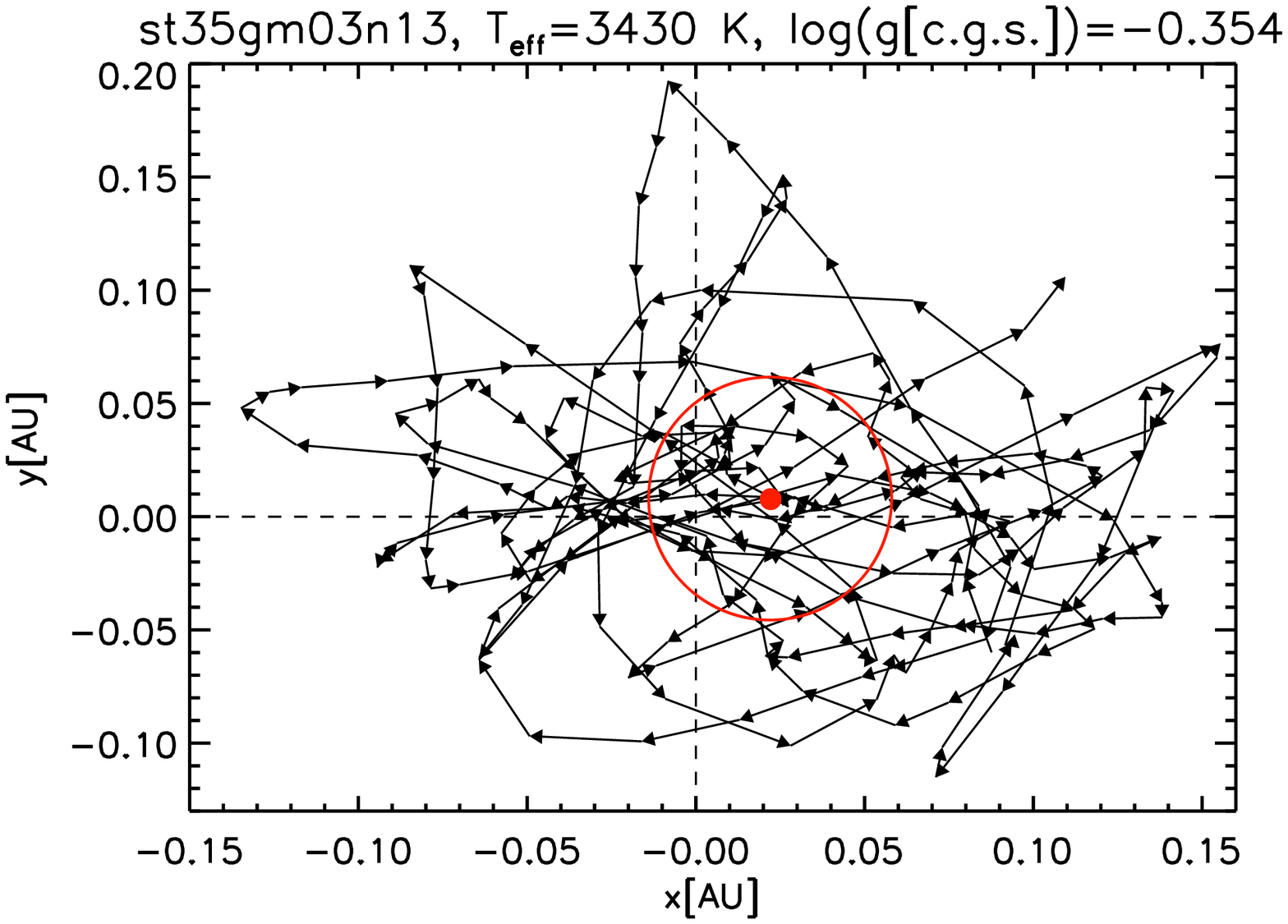} 
          \hspace{0.8cm}
       \includegraphics[width=0.47\hsize]{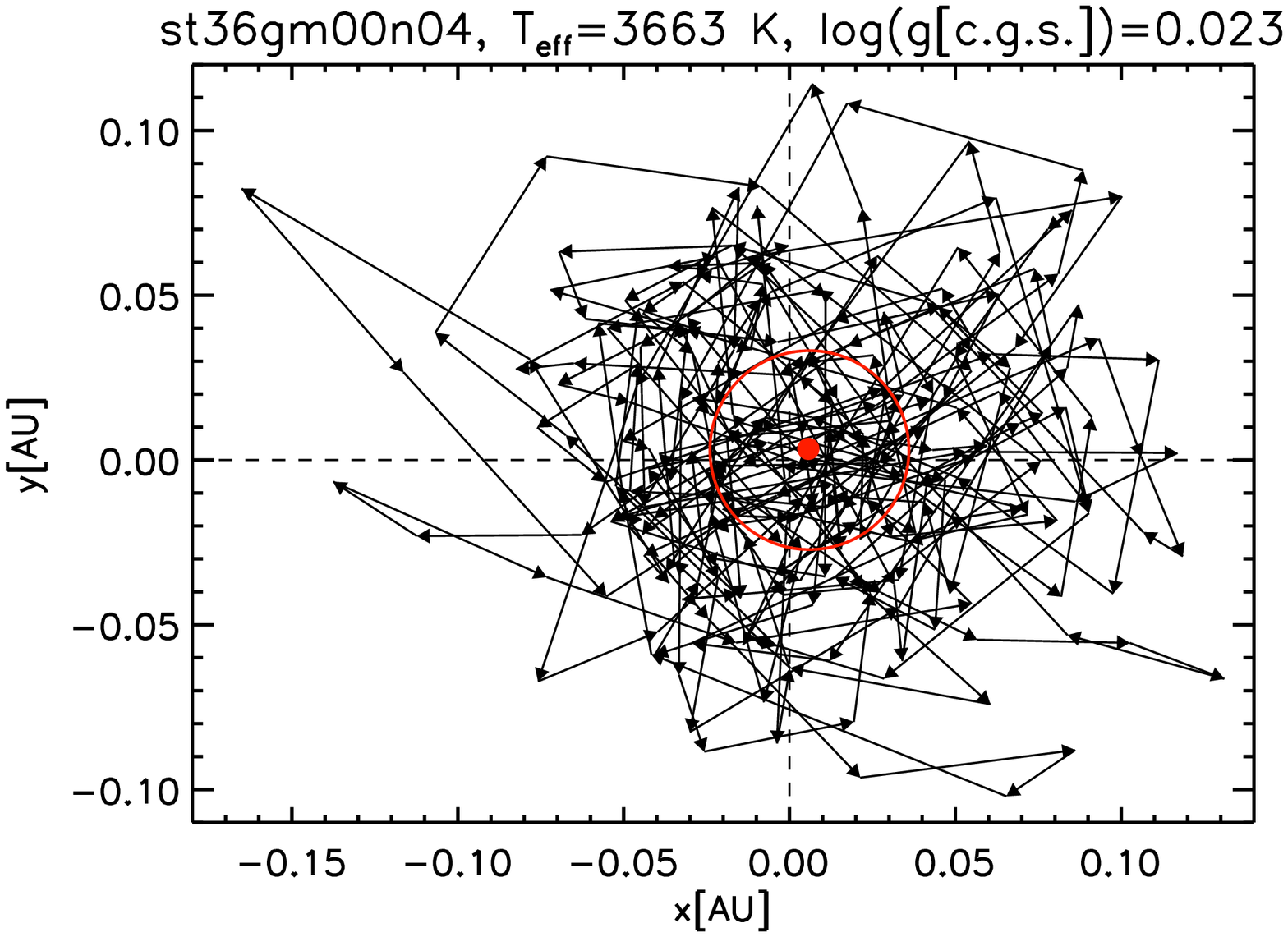} \\
        \includegraphics[width=0.47\hsize]{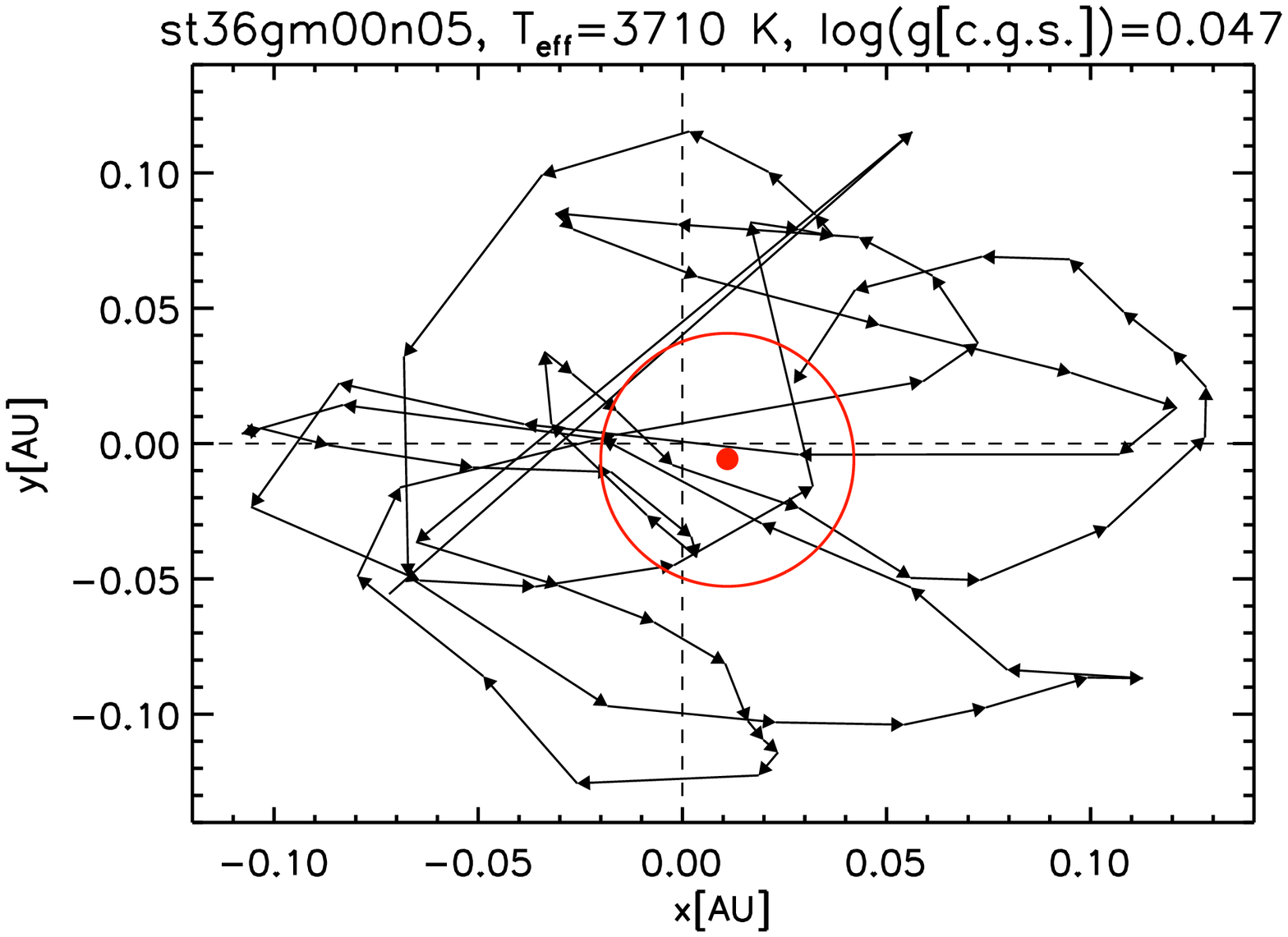} 
            \hspace{0.7cm}
         \includegraphics[width=0.47\hsize]{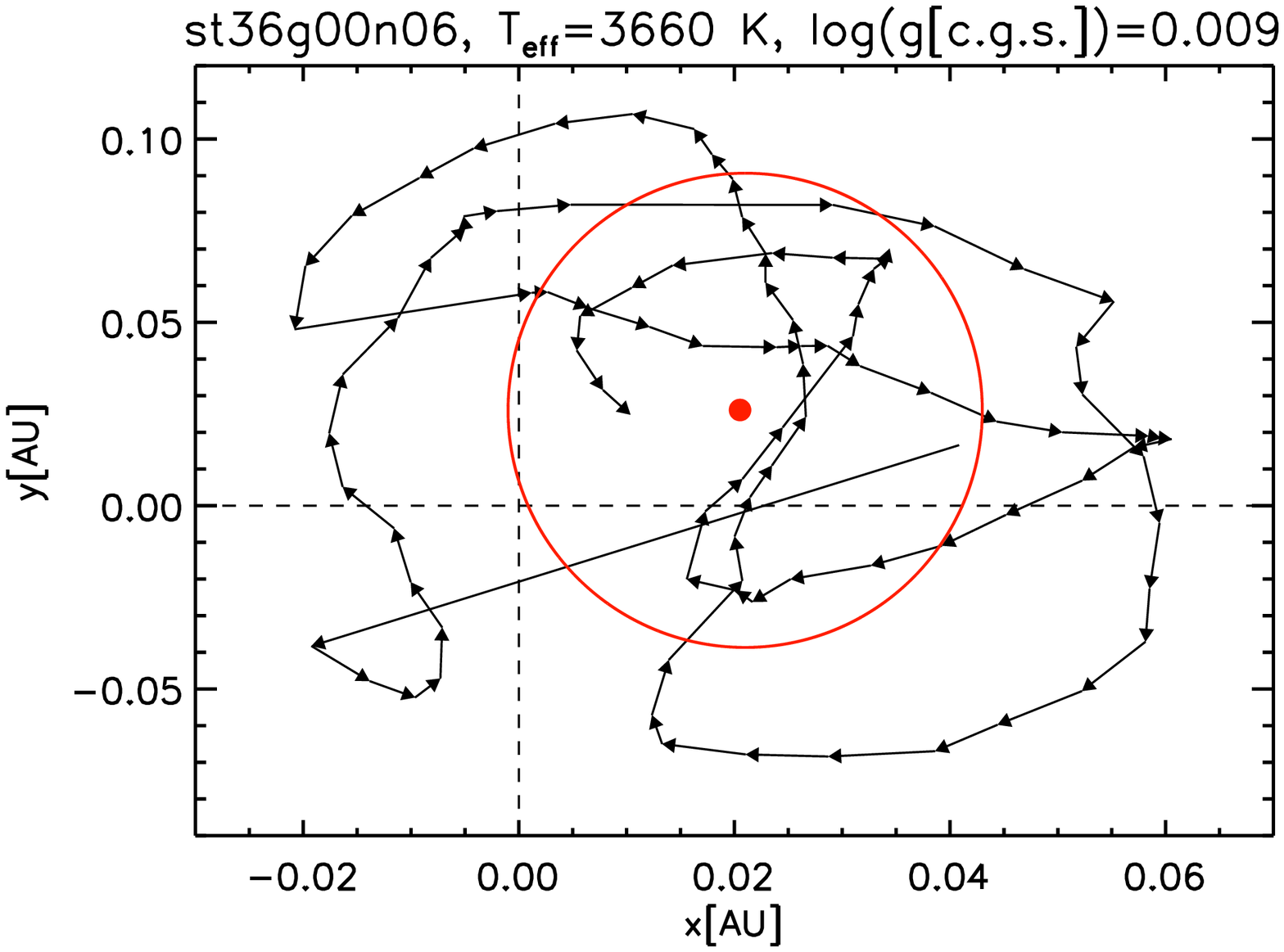} 
  \end{tabular}
\caption{Photo-center positions computed from the intensity maps of the RHD simulations in Table~\ref{simus} in the Gaia $G$ photometric system. The different snapshots are connected by the line segments, the total time covered is reported in the Table. The time interval between two consecutive points is $2\times10^6$\,s (about 23 days) for all simulations except for st35gm04b1n001, for which it is $4\times10^6$\,s. The dashed lines intersect at the position of the geometrical center of the images while the red dot and the red circles display the expected observable position of the star $\langle P\rangle$ with $\sigma_P$ uncertainty.} 
\label{photo_appendix}
\end{figure*}
  
%\begin{figure}%f1
%\includegraphics[width=0.95\hsize]{images/dst28gm07n001/photocenter.ps}
%\caption{Same as in Fig.~\ref{photo1} for RHD simulation st28gm07n001 in Table~\ref{simulations}.} 
%\label{photo7}
%\end{figure}

\end{appendix}

\end{document}